\documentclass[conference]{IEEEtran}
\IEEEoverridecommandlockouts
\usepackage{url}

\usepackage{breakurl}
\usepackage[breaklinks]{hyperref}
\usepackage{cite}
\usepackage{amsmath,amssymb,amsfonts}
\usepackage{algorithmic}
\usepackage{graphicx}
\usepackage{textcomp}
\usepackage{xcolor}
\usepackage{balance}
\def\BibTeX{{\rm B\kern-.05em{\sc i\kern-.025em b}\kern-.08em
    T\kern-.1667em\lower.7ex\hbox{E}\kern-.125emX}}
\begin{document}

\title{Secure Cloud Storage with Client-Side Encryption Using a Trusted Execution Environment}

\author{\IEEEauthorblockN{Marciano da Rocha}
\IEEEauthorblockA{\textit{Department of Computer Science} \\
\textit{Federal University of Technology - Paraná}\\
Dois Vizinhos, Brazil \\
marciano@alunos.utfpr.edu.br}
\and
\IEEEauthorblockN{Dalton Cézane Gomes Valadares}
\IEEEauthorblockA{\textit{Department of Mechanical Engineering} \\
\textit{Federal Institute of Pernambuco}\\
Caruaru, Brazil \\
dalton.valadares@caruaru.ifpe.edu.br}
\and
\IEEEauthorblockN{Angelo Perkusich}
\IEEEauthorblockA{\textit{Department of Electrical Engineering} \\
\textit{Federal University of Campina Grande}\\
Campina Grande, Brazil \\
perkusic@dee.ufcg.edu.br}
\and
\IEEEauthorblockN{Kyller Costa Gorgonio}
\IEEEauthorblockA{\textit{Department of Computer Science} \\
\textit{Federal University of Campina Grande}\\
Campina Grande, Brazil \\
kyller@dsc.ufcg.edu.br}
\and
\IEEEauthorblockN{Rodrigo Tomaz Pagno}
\IEEEauthorblockA{\textit{Department of Computer Science} \\
\textit{Federal University of Technology - Paraná}\\
Dois Vizinhos, Brazil \\
rodrigopagno@utfpr.edu.br}
\and
\IEEEauthorblockN{Newton Carlos Will}
\IEEEauthorblockA{\textit{Department of Computer Science} \\
\textit{Federal University of Technology - Paraná}\\
Dois Vizinhos, Brazil \\
will@utfpr.edu.br}
}

\maketitle

\begin{abstract}
With the evolution of computer systems, the amount of sensitive data to be stored as well as the number of threats on these data grow up, making the data confidentiality increasingly important to computer users. Currently, with devices always connected to the Internet, the use of cloud data storage services has become practical and common, allowing quick access to such data wherever the user is. Such practicality brings with it a concern, precisely the confidentiality of the data which is delivered to third parties for storage. In the home environment, disk encryption tools have gained special attention from users, being used on personal computers and also having native options in some smartphone operating systems. The present work uses the data sealing, feature provided by the Intel Software Guard Extensions (Intel SGX) technology, for file encryption. A virtual file system is created in which applications can store their data, keeping the security guarantees provided by the Intel SGX technology, before send the data to a storage provider. This way, even if the storage provider is compromised, the data are safe. To validate the proposal, the Cryptomator software, which is a free client-side encryption tool for cloud files, was integrated with an Intel SGX application (enclave) for data sealing. The results demonstrate that the solution is feasible, in terms of performance and security, and can be expanded and refined for practical use and integration with cloud synchronization services.
\end{abstract}


\section{Introduction}
\label{sec:Intro}

Cyber attacks and different cybercrimes are an increasing threat to today's digital society \cite{Huang2018,Khraisat2019,Singh2019}. In this context data governance and security is of utmost importance \cite{Onwujekwe2019}.  Reports from specialized companies and government security entities indicate a growing number of threats to digital data, whether from corporations or users. Companies faced an average of 25 new threats per day in 2006, up from 500,000 in 2016 \cite{Weafer2016} and 61\% of CEOs are concerned with the state of the cyber security of their company \cite{PWC2016}. 

Nowadays, users are increasingly using cloud storage services to keep their files and can access them from any device connected to the Internet. Such storage services hold a large set of data from various users, including sensitive and confidential information, due to their reliance on such providers. However, these storage services do not always guarantee or respect the privacy of their users \cite{Branscombe2015,Cox2016,Clover2017}.

One way to ensure an additional level of security for users' sensitive files is to encrypt such data before sending them to the cloud storage server. One option that performs such an operation is using a disk encryption system, which runs in the operating system background and encrypts all information that is stored.

This type of system has some vulnerabilities, which can highlight the risk of attacks and improper access to data. One of these vulnerabilities is the user's password choice, which is often weak and can be discovered using dictionary attacks, default passwords, rainbow tables or even brute force. Another factor is that most of the existing disk encryption systems keep the encryption key in main memory while it is running, allowing other malicious applications to attack and discover the key.

Nowadays the use of reliable trusted hardware technologies, such as Intel Software Guard Extensions (Intel SGX) \cite{IntelSGXProgRef} and  ARM TrustZone \cite{Pinto2019}, has grown. An Intel SGX application creates an isolated protected region of memory, which is called enclave, where data are securely processed. Among the various features provided by the SGX technology, there is the sealing of an enclave data in secondary memory. This process uses a unique key as a basis for the encryption, which is generated by combining a developer key and a processor key, and is not stored in memory: it is generated again to each request, guaranteeing that the data will be decrypted only by the enclave that sealed them, in the platform in which they were sealed \cite{Anati2013}. Another feature provided by Intel SGX technology is the encryption of data in a region of primary memory with a random key that is generated with each power cycle. This key is known only by the processor and never goes beyond its limits \cite{Intel2016}.

According to the previous discussion, we established the following two research questions to guide our research:
\begin{enumerate}
\item How to improve the security of traditional disk encryption tools? 
\item How to decrease the user's need to trust the cloud service provider?
\end{enumerate}

In this paper, we propose the use of Intel SGX features to perform data sealing and processing in a more secure way, since these operations are performed only inside an enclave, which is supposed to be protected even against adversaries with high privileges inside the operating system. We implemented, as a proof of concept, an application using the Cryptomator, an open source client-side encryption tool used for cloud files protection. The Cryptomator was modified, integrated with an Intel SGX application, in order to use its data sealing feature, which is performed only inside an enclave. With this, the data encryption/decryption processes are performed in a more secure way, extending this security also for the keys, since they are also handled inside the enclave.

The main contributions of this work are listed below:
\begin{itemize}
\item The proposal to improve the security of disk/file encryption process using Intel SGX;
\item The implementation of such proposal considering an Intel SGX application integrated with the Cryptomator tool;
\item The performance evaluation of our proposal implemented, considering six different hardware combinations.
\end{itemize}

The remainder of this paper is organized as follows: Section \ref{sec:Background} provides background information about the cloud storage and disk encryption techniques, as well the Intel Software Guard Extensions technology; Section \ref{sec:RelatedWork} presents previous research closely related to our proposal, in the field of cloud storage security and the use of Intel SGX in file encryption; Section \ref{sec:Proposal} describes the solution proposed by this paper; in order to validate the proposed approach, three prototypes are presented in Section \ref{sec:Implementation} as a proof of concept; Section \ref{sec:Performance} presents the performance evaluation of the prototype on six different hardware combinations; the threat model as well the security evaluation of the Intel SGX technology and the proposal solution are presented in Section \ref{sec:Security}; the limitations of our solution are described in Section \ref{sec:Limitations} and, finally, Section \ref{sec:Conclusion} concludes the paper and presents the future works.

\section{Background}
\label{sec:Background}

This Section presents a brief description of important concepts used in this work, such as cloud storage, disk encryption and the Intel Software Guard Extensions.

\subsection{Cloud Storage}

Nowadays, more users are using cloud storage services, either explicitly to keep their personal files, or implicit through applications that make use of such background services to maintain backups or history of user data. The benefits inherent in cloud data storage are virtually unlimited storage, version history for each file, and access to data at any time from any device connected to the Internet. 

But there is always a concern regarding the security of data stored in the cloud, especially as the number of constant cyber attacks intensifies. One of the most promising defenses for the cloud is encryption, which offers robust protection in both data storage and transit. Encryption is available in two main security configurations: client- and server-side.

With server-side encryption, the cloud storage provider manages encryption keys along with their data, and many cloud storage providers use this method. This limits the complexity of the environment while maintaining the isolation of your data, but there are several risks of keeping the encryption key and data encrypted by it in the same place. In addition, the cloud provider itself can access keys and hence data in plain text.

In client-side encryption, the user is responsible for keeping their keys, accessed with a password, which is a user-centric approach to keep control of the data. This configuration limits the risk of external access to sensitive information, by unauthorized entities, since encryption is ideal for confidential data. Most client-side encryption methods involve encrypting data for storage in a cloud service that does not support natively. 

The user can determine which data to encrypt and which data to store in the cloud in plain text. The user can encrypt only certain files, or create an entire encrypted file container or storage folder. This can be done even when using server-side encryption services, adding an extra layer of security to extremely sensitive data.

\subsection{Disk Encryption}

Many people and companies have sensitive information stored on the most diverse types of devices. In this context, Full Disk Encryption (FDE) has emerged as a solution to ensure the security of these information in cases of theft or loss of devices that store them, since mechanisms such as passwords and biometric blocking do not prevent the storage media from being accessed when installed on another machines. FDE is an effective method of protecting data from unauthorized access, because it consists of encrypting entire volumes/partitions or disks, ensuring that information can not be accessed without the mechanism and key used for encryption. Existing mechanisms can be classified into firmware-level encryption (hardware-based solutions) and kernel-level encryption (software-based solutions).

Hardware-based FDE engines are called Self-Encrypting Drives (SEDs), in which encryption is performed by the disk controller and the encryption keys are not present in the CPU neither in the main memory of the computer. In addition, the MBR (Master Boot Record) is also encrypted, preventing manipulation attacks. This type of FDE is present in some specific storage media models, such as Intel SSD 320 and 520 drives, and requires a previous boot environment with a screen available for entering the access password. According to \cite{SSDSecurityPaper}, some Solid State Drive (SSD) models have failures in the encryption process, allowing the stored data to be extracted from the drive without knowing the key used to perform the encryption.

Software-based FDE engines are applications run at the kernel level by the CPU, in which data and used keys are stored in the main memory. This type of FDE consists of intercepting all operating system (OS) requests and encrypting the data before storage, or fetching encrypted data on storage media and returning OS-readable information. Software-based FDE do not encrypt the MBR, which allows manipulative attacks on it. This kind of solution has been on the market since the 2000s, beginning with the launch of \emph{Cryptoloop} \cite{Cryptoloop} for Linux, which was the predecessor of \emph{Dm-Crypt} \cite{DMCrypt}, released in 2003.

Linux based operating systems commonly implement the LUKS (Linux Unified Key Setup) disk-encryption specification. LUKS is a platform-independent standard on-disk format based on a two level key hierarchy, protecting the master key using PBKDF2 \cite{RFC2898} as key derivation function, with an anti-forensic splitter (AF-splitter) to solve the problem of data remanence, that inflates and splits the master key before storing it on disk \cite{luks,Bossi2015}.

In addition to these, other disk encryption systems are also widely used by end users, such as Cryptomator, that uses a virtual file system to create a file container, where the encrypted data are stored \cite{Cryptomator}.

\subsection{Intel SGX}
\label{sec:SGX}

Intel SGX technology compiles a set of instructions and mechanisms for creating and accessing a protected region of memory, ensuring the confidentiality and integrity of sensitive application data. The first version of SGX was added to Intel Core processors from the 6th generation (Skylake) and allows an application to start a protected container, called \emph{enclave}. In this context, two significant features have been added in Intel's x86 architecture: a change in the enclave memory access and the protection of application address mappings \cite{IntelSGXProgRef,McKeen2013}.

An enclave is a fixed-size protected area in the application's address space, which allows a portion of the application code to be run confidentially and securely, ensuring that other software can not access these information, even if it has high running privilege or if it is running within other enclaves \cite{IntelSGXProgRef}. Attempts to gain unauthorized access to the contents of an enclave are detected and prevented, or the operation is aborted. While the enclave data are being processed between the registers and other internal blocks of the processor, it uses internal access control mechanisms to prevent unauthorized access to these data. When data are transferred to main memory, they are automatically encrypted and stored in a reserved region, called Processor Reserved Memory (PRM).

Memory encryption is done using a 128-bit time-invariant AES-CTR encryption algorithm and containing protections against replay attacks. The encryption key is stored in internal registers of the processor, not being accessible to external components, and is randomly changed at each hibernation or system restart event. Memory probes or other techniques that attempt to modify or replace these data are also avoided, and the fact of connecting the memory module to another system will only give access to the data in an encrypted form \cite{Aumasson2016,IntelLinuxSDK}.

Intel SGX technology ensures the confidentiality and integrity of the data while inside an enclave. In general, when the enclave is destroyed, all information is lost. To preserve these information, SGX technology provides a mechanism that allows the enclave to seal them using a key called \emph{Sealing Key}, so the information can be safely stored on disk. The sealing key is a unique key generated by the CPU for that enclave and on that particular platform, and it is not necessary to store it for unsealing the data. Sealing and unsealing are the operations involved in the data protection process on disk \cite{Intel2016}. The AES-GCM algorithm is used for data sealing and the AES-CMAC algorithm is used for key derivation \cite{Aumasson2016}.

In order to share information, protected in a running enclave, with other enclaves, the security must be ensured at the destination and also during the transmission process. To this end, Intel SGX technology provides a feature called attestation, which allows an enclave to prove to third parties that it is legitimate, unadulterated and correctly loaded, allowing the creation of a secure channel for communication between them. Intel SGX technology provides mechanisms that enable two forms of attestation: local attestation and remote attestation \cite{Intel2016}.

Local attestation is performed when two enclaves on the same device need to securely exchange information with each other. In this case, one enclave must prove its identity and authenticity to the other in order to begin the communication. This form of attestation uses a symmetric key system,
with the enclave asking the hardware a credential, which should be forwarded to the other enclave who will verify that the credential was generated on the same platform. The used key is embedded in the hardware platform, and only known by the enclaves running on that platform.

Remote attestation is performed when enclaves are running on separate devices and need to securely exchange information with each other or when a third party needs to securely send data to an enclave. This form of attestation requires the use of asymmetric encryption, a special Intel-provided enclave called the \emph{Quoting Enclave}, and a CPU-produced credential called \emph{QUOTE}. The QUOTE credential is generated by replacing the Message Authentication Code (MAC) of the reports generated by local enclaves with a signature created using an asymmetric key using the Intel Enhanced Privacy ID (EPID). To ensure the security of the Intel EPID key, only the instantiated Quoting Enclave has access to it \cite{Anati2013,Intel2016}.

\section{Related Work}
\label{sec:RelatedWork}

This Section contains related works that also consider cloud storage security and privacy, and disk encryption using the Intel SGX technology.

\subsection{Cloud Storage Security and Privacy}

The security and privacy of files stored in the cloud are a central concern because such data are being delivered to third parties. Several works in the literature present different techniques and mechanisms to guarantee or increase the security of users' files and maintain their confidentiality.

One of the features widely used by cloud storage providers to improve storage utilization is deduplication, which aims to eliminate duplicate copies of repeating data. Deduplication can also optimize the network data traffic by reducing number of bytes that must be sent, but can also be used as a side channel technique by attackers who try to obtain sensitive information of other users' data. In order to mitigate this vulnerability, client-side encryption schemes are proposed that allow data deduplication and audit while prevent leakages \cite{Shin2014,Youn2018}.

Audit operations can also characterize a point of failure, allowing encryption key exposure. To deal with this problem, in \cite{Yu2017}, the authors point out that cloud storage auditing scheme with key-exposure resilience has been proposed, but valid authenticators can still be forged later than the key-exposure time period, if the current secret key of the data owner has been obtained. Then, the authors propose a new kind of cloud storage auditing and design a concrete scheme where the key exposure in one time period does not affect the security of cloud storage auditing in other time periods.

A relationship between secure cloud storage and secure network coding is also demonstrated by \cite{Chen2016}, what originates a systematic way to construct secure cloud storage protocols based on any secure network coding protocol. The authors also propose a publicly verifiable secure cloud storage protocol that support user anonymity and third-party public auditing.

Also, \cite{Gopinaath2017} presents a server-side encryption approach with federation sharing for cloud storage using a hybrid environment. An approach to split the data and distribute them along different cloud servers is described in \cite{Li2017}, to prevent cloud service operators from directly accessing partial data. The authors also present an alternative approach to determine whether the data packets need a split to shorten the operation time.

Hardware-based security mechanisms are also used to provide security and privacy in cloud storage services. In \cite{Crocker2015}, the authors describe a two factor encryption architecture that incorporates the use of a hardware token. A proof of concept was developed with a middleware that can be used with any cloud storage provider that makes use of the OAuth 2.0 protocol for authentication and authorization.

In \cite{Valadares2018}, Trusted Execution Environments (TEEs) are proposed to increase the security and privacy of data storage and processing in cloud/fog-based IoT applications. The authors propose an architecture that applies authentication and authorization for the participants, and cryptography for the generated data. These data must be decrypted and processed only inside a TEE application. A proof of concept using Intel SGX was implemented and evaluated, presenting an acceptable communication latency when compared to an application that did not apply any security mechanism.

\subsection{Disk Encryption with Intel SGX}

Intel SGX technology has been used in a wide range of applications and in several areas. One of these areas is disk encryption, in which Intel SGX provides an additional layer of security for the storage of sensitive files.

In \cite{Richter2016} the authors introduce the concept of isolating kernel components from the operating system into enclaves in order to prevent vulnerabilities in certain modules from completely compromising the system. Due to the restrictions of the Intel SGX technology, the enclave can not be executed directly by the kernel, so the authors had to include a daemon running in user mode to communicate with a Loadable Kernel Module (LKM). As proof of concept, the authors created an LKM that registers a new mode inside the kernel encryption API (Application Programming Interface), allowing to carry out the encryption process within an enclave, that can be used for disk encryption.

The proposal presented by \cite{Burihabwa2018} is to include the data encryption within a file system based on the FUSE\footnote{https://github.com/libfuse/libfuse} (Filesystem in Userspace), using the Intel SGX technology to ensure that the stored data are secure. File requests made to the operating system through the virtual file system are intercepted by the FUSE library and sent to an enclave, which performs the process of encrypting and decrypting the data using the native data sealing feature provided by the Intel SGX technology.

The paper \cite{Ahmad2018} focuses on decreasing the chances of success in a side channel attack on a file system based on the Intel SGX technology. It consists of a library that has a file system running within an enclave, and an application running in another enclave making requests to that file system. Communication between the enclave library and the enclave application is performed through a queue of messages sent by encrypted communication channels between the processes.

\section{Client-Side Encryption using TEE}
\label{sec:Proposal}

This work proposes the inclusion of the data sealing feature, provided by the Intel SGX technology, integrated to a Disk/File Encryption Tool. In order to validate our proposal, we are considering the Cryptomator software. Cryptomator works with file container encryption, providing the user's operating system with a virtual file system where data are read and written elsewhere.

Cryptomator is designed to be used as a client-side file encryption service and can work with any cloud data synchronization software \cite{Horalek2018}. In addition, Cryptomator provides the user with a transparent service by encrypting the files individually, allowing the cloud storage service to maintain a file update history. Another feature provided by Cryptomator is directory structure obfuscation \cite{Cryptomator}.

The application is divided into three main modules, namely:
\begin{itemize}
    \item \textbf{Cryptomator:} graphical interface that provides the user with control of the containers;
    \item \textbf{CryptoFS:} library that implements a virtual file system and is responsible for reading and writing data within the containers, through the operating system, providing the file system with decrypted data and encrypting the received data before storing them on a secondary media;
    \item \textbf{CryptoLib:} library that provides functions for the encryption and decryption of the files, which are handled by the CryptoFS module.
\end{itemize}

\begin{figure}[ht]
    \centering
    \includegraphics[width=\columnwidth]{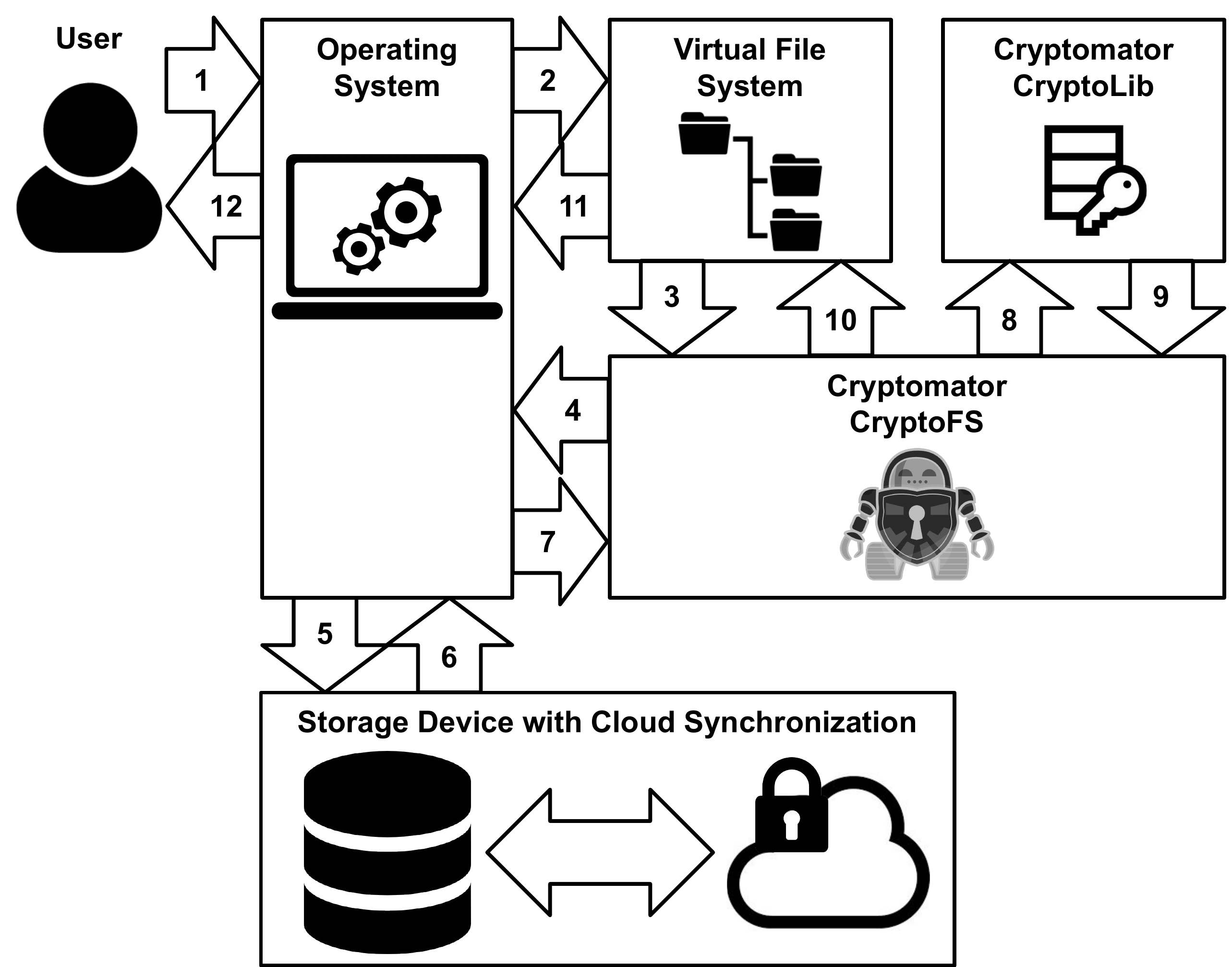}
    \caption{Workflow for reading and decrypting stored data with Cryptomator.}
    \label{fig:cryptomator}
\end{figure}

The process of reading the files provided by the application can be described in 12 steps, as shown in Fig. \ref{fig:cryptomator}. These steps are explained below:
\begin{enumerate}
    \item The user requests the operating system to open a file;
    \item The operating system requests FUSE for file data. FUSE allows the userspace applications export a filesystem to the Linux kernel, with functions to mount the file system, unmount it and communicate with kernel;
    \item FUSE forwards this request to the Cryptomator, using the CryptoFS library;
    \item Cryptomator requests the operating system to have the file data fetched from the storage device;
    \item The operating system locates the data;
    \item These data are loaded into the main memory;
    \item The operating system provides these data to the Cryptomator;
    \item The CryptoFS library sends the encrypted data to the CryptoLib library;
    \item The CryptoLib library decrypts the received data and returns them to CryptoFS library;
    \item The CryptoFS library sends the decrypted data to FUSE;
    \item FUSE forwards such data to the operating system;
    \item Finally, the operating system provides the user with the decrypted file.
\end{enumerate}
    
The data writing process is similar to the reading process, but with the plain data being sent to the CryptoLib library before, and then they are forwarded to the operating system for recording in the storage device. The files stored in the local device are also synchronized with a cloud storage service, already in encrypted form, and a second layer of encryption can be applied by the cloud storage provider.

In this sense, we propose to include the data sealing mechanism provided by Intel SGX technology within the CryptoLib library, in parallel with the existing AES encryption, and to change the CryptoFS library to use this modified implementation. The proposed change ensures, through the use of existing AES encryption, that the CryptoLib library can still be used in environments where Intel SGX technology is not available, thereby maintaining project compatibility with changes in the main design.

\section{Proof of Concept}
\label{sec:Implementation}

Cryptomator is designed to support multiple encryption modes, without any of them affecting the implementation of the others or requiring changes to the main project to support them. The CryptoLib library is composed of a Java interface (\texttt{API}), which describes all the classes and methods that each encryption mode must implement to be used by the software, and an implementation of the AES encryption based on the described interface, which is called \texttt{v1} within the library.

Our implementation consists of creating a new mode called \texttt{sgx}, implementing all classes and methods requested by the library interface, and changing the CryptoFS library to use \texttt{sgx} mode instead of \texttt{v1}. It was also necessary to create a C++ library for communication with the SGX enclaves, called \texttt{SgxLib}.

The \texttt{sgx} mode has a Java class with 4 main methods used by all other classes of the mode, namely:
\begin{itemize}
    \item \textbf{InitializeEnclave}: handles the initialization of the SGX enclave;
    \item \textbf{SgxEncryptBytes}: encrypts an array of bytes;
    \item \textbf{SgxDecryptBytes}: decrypts an array of bytes;
    \item \textbf{DestroyEnclave}: destroys the enclave when it is no longer needed.
\end{itemize}

These 4 methods are only used to invoke the corresponding methods within the \texttt{SgxLib} library. In order to handle the communication between Java methods and the \texttt{SgxLib} library, it was necessary to create a Java Native Interface (JNI), containing the 4 main methods of the Java class. This interface has the purpose of receiving the Java application data, converting them to the C++ language, performing the process implemented by the method and then converting its results and sending them to Java.

\begin{figure}[ht]
    \centering
    \includegraphics[width=\columnwidth]{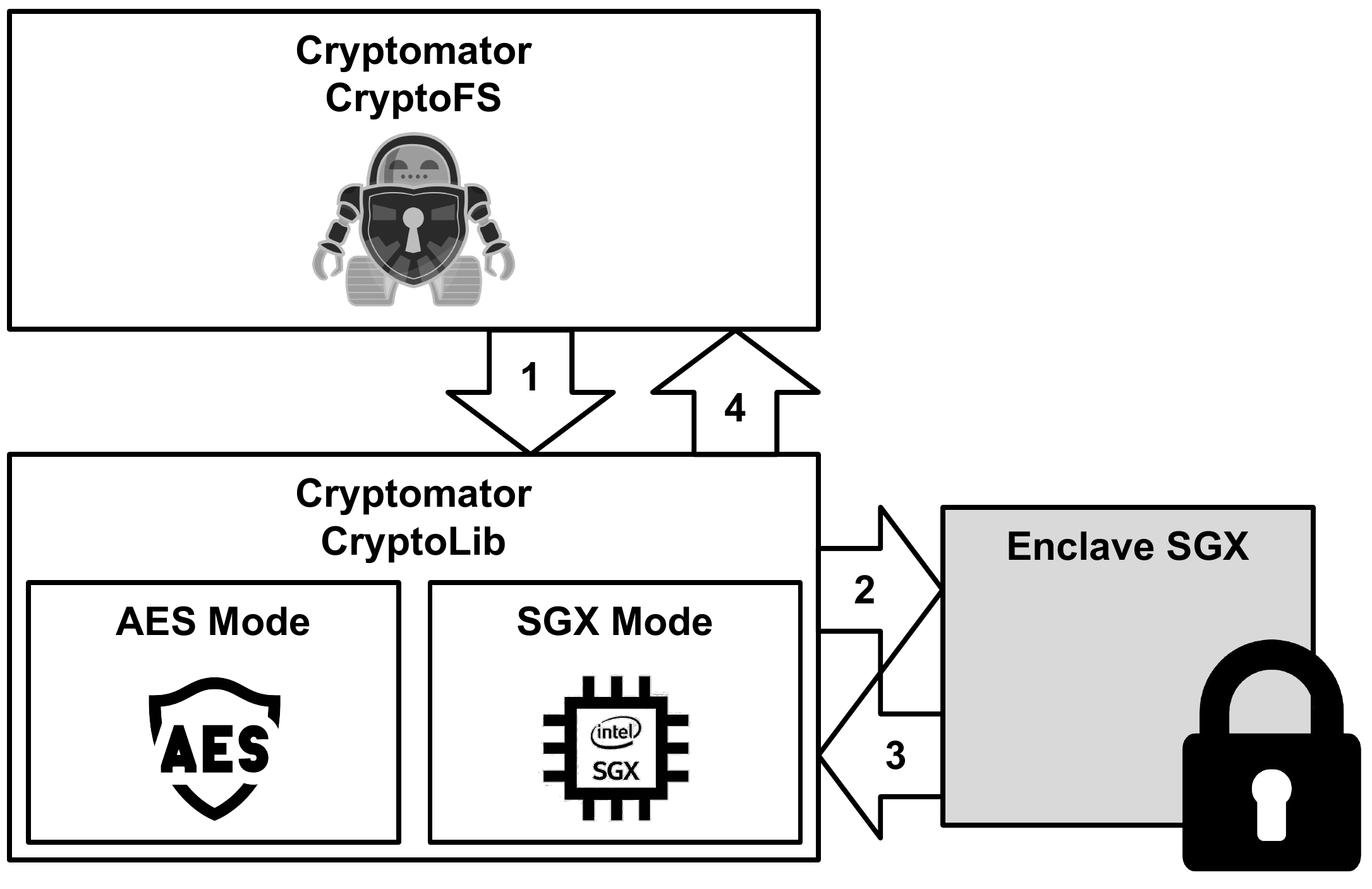}
    \caption{Workflow performed for data encryption and decryption through the Cryptomator with the inclusion of data sealing feature provided by Intel SGX technology.}
    \label{fig:cryptomatorsgx}
\end{figure}

Fig. \ref{fig:cryptomatorsgx} describes the process of sending and receiving data between CryptoFS and CryptoLib libraries, which includes two additional steps for communication between the CryptoLib library and the SGX enclave. The shaded box indicates a trusted execution environment (SGX in this case), where the key is manipulated and the data are encrypted and decrypted. The communication flow is explained below:
\begin{enumerate}
    \item CryptoFS library calls encrypt/decrypt functions in CryptoLib library, which contais AES and SGX modes;
    \item In SGX mode, CryptoLib library creates an enclave and sends to it each data block to perform the encrypt/decrypt process;
    \item Encrypt/decrypt blocks are send back to CryptoLib;
    \item Encrypt/decrypt data forward to CryptoFS library.
\end{enumerate}

\section{Performance Evaluation}
\label{sec:Performance}

To evaluate the proposal based on the implemented proof of concept, performance tests were carried out comparing four modes of data storage, and are as follows:
\begin{itemize}
    \item \textbf{Without Encryption:} Data read and write operations on storage media without any type of encryption;
    \item \textbf{LUKS:} Data read and write operations on storage media using native cryptographic mode in Ubuntu with LUKS (Linux Unified Key Setup);
    \item \textbf{Cryptomator:} Data read and write operations on the storage media using the original Cryptomator application;
    \item \textbf{Cryptomator-SGX:} Data read and write operations on the storage media using the implemented solution, which integrates the SGX data sealing with the Cryptomator application.
\end{itemize}

For each mode described above, four different tests were performed, two for data read operations and two for data write operations. The tests considered the transferring of a single file and also multiple files in sequence. For the task, the DVD ISO image of the CentOS 7 operating system \cite{CentOS}, which has a size of 4.27 GB, was used as the data set for the single file transfer, as well as the files and folders obtained when extracting the same image, using the recursive transfer of directories, were used characterizing the transfer of several files.

The \emph{RSync} \cite{rsync} tool was used to perform data transfer, using \texttt{-avhh} modes for single file and \texttt{-rvhh} for multiple files (directories as execution parameters). Further details on the parameters used are outlined below:
\begin{itemize}
    \item \textbf{-a}: Single file copy mode;
    \item \textbf{-r}: Recursive copy mode of directories, in which all sub-directories and files are copied too;
    \item \textbf{-v}: Option to print what is running at the command prompt;
    \item \textbf{-h}: Mode to display a better format to read numbers. This causes large numbers to avoid using larger units, considering a suffix K, M, or G. With this option specified once, these units are K (1000), M (1000 * 1000) and G (1000 * 1000 * 1000); with this option repeated, the units are powers of 1024 instead of 1000.
\end{itemize}

\subsection{Experimental Setup}

Tests were run in custom PC, motherboard with a Z390 chipset, 16 GB 2666 MHz RAM, SGX enabled with 128 MB PRM size, running Ubuntu 16.04.6 LTS, kernel 4.15.0-51-generic. Intel TurboBoost, SpeedStep, and HyperThread extensions were disabled, to get stable results. We used the Intel SGX SDK 1.7 for Linux and set the stack size at 4 KB and the heap size at 1 MB.

In order to measure the CPU contribution to the performance impact of the encryption task, two distinct CPUs were used:
\begin{itemize}
    \item \textbf{CPU 1:} Dual-core 3.8 GHz Intel Pentium G5500;
    \item \textbf{CPU 2:} Octa-core 3.6 GHz Intel i7 9700K.
\end{itemize}

We also used three distinct storage devices to perform the tests, each one containing different characteristics regarding performance for data reading and writing:
\begin{itemize}
    \item \textbf{Device 1:} HDD Samsung ST1000LM024, 1 TB storage size, 5400 RPM spin speed, 145 MB/s maximum data transfer rate, 8 MB buffer DRAM size;
    
    \item \textbf{Device 2:} SSD SanDisk PLUS, 240 GB storage size, 530 MB/s sequencial read, 440 MB/s sequencial write;
    
    \item \textbf{Device 3:} SSD NVMe M.2 Samsung 970 EVO Plus, 250 GB storage size, 3500 MB/s sequencial read, 2300 MB/s sequencial write.
\end{itemize}

In all devices, an Ext4 file system partition was used. Since one of the storage devices has high performance (SSD M.2), and the highest performance in a data transfer between two devices is defined by the medium with lower performance, a \emph{RAMDisk} was used to perform the recording tests, as data source and target for the read tests. This way, there was no speed limitation by the source or destination of the data. 

Each test was performed 10 times, aiming to obtain the average transfer rate in each scenario considered. Another set of tests was also performed to analyse the reading performance of data previously stored in each of the devices. The single file reading and a set of different files were tested.

\subsection{Results and Discussion}

\begin{figure*}[ht]
    \centering
    \includegraphics[width=.9\textwidth]{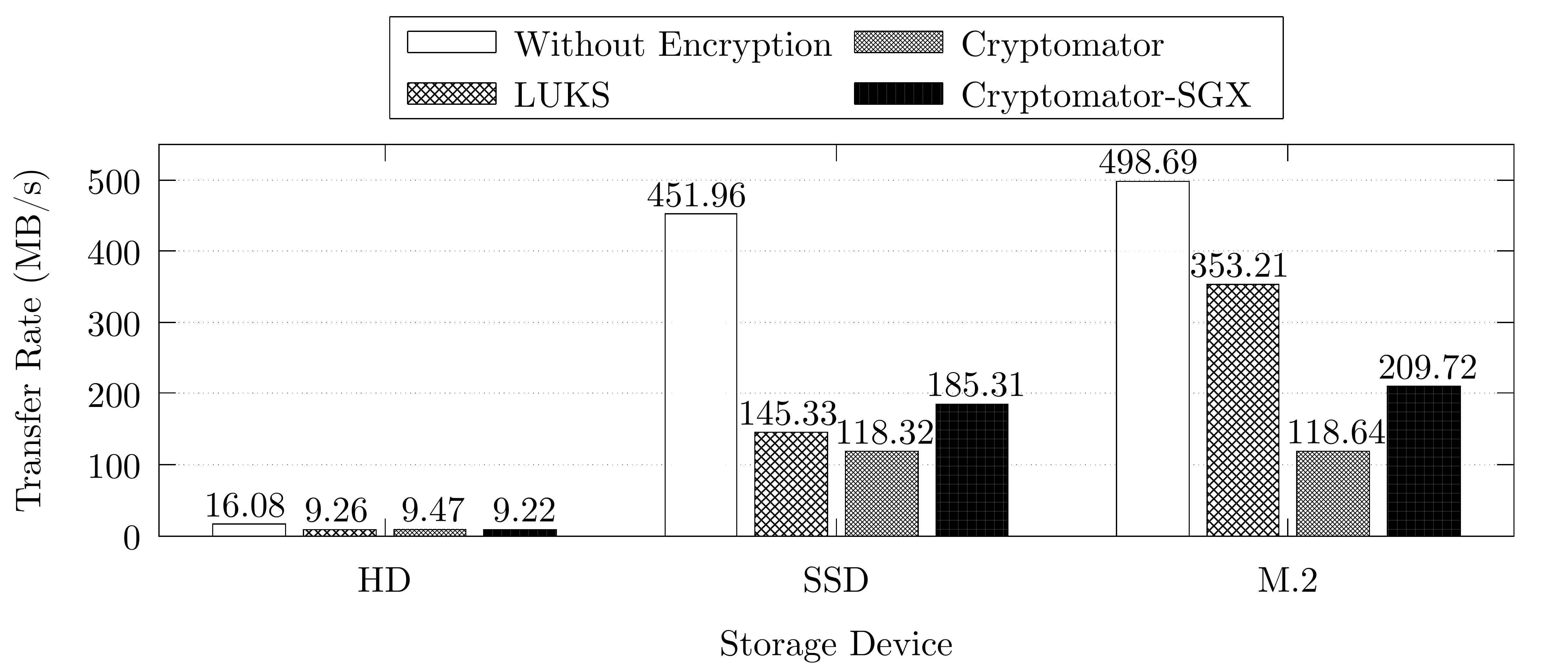}
    \caption{Transfer rate reading a single file with Intel Pentium G5500 CPU.}
    \label{fig:pentium_read_iso}
\end{figure*}

The results obtained in the single file read, on the three storage devices using the Intel Pentium G5500 CPU, can be seen in Fig. \ref{fig:pentium_read_iso}. The performance of Cryptomator-SGX solution is higher than its original implementation, with the exception of the hard disk, where the throughput was slightly below that observed in the unmodified Cryptomator implementation, but still very close to that. It is also observed that, in this scenario, the proposed solution obtained transfer rates very close to or even above those presented by LUKS, except in storage device M.2.

\begin{figure*}[ht]
    \centering
    \includegraphics[width=.9\textwidth]{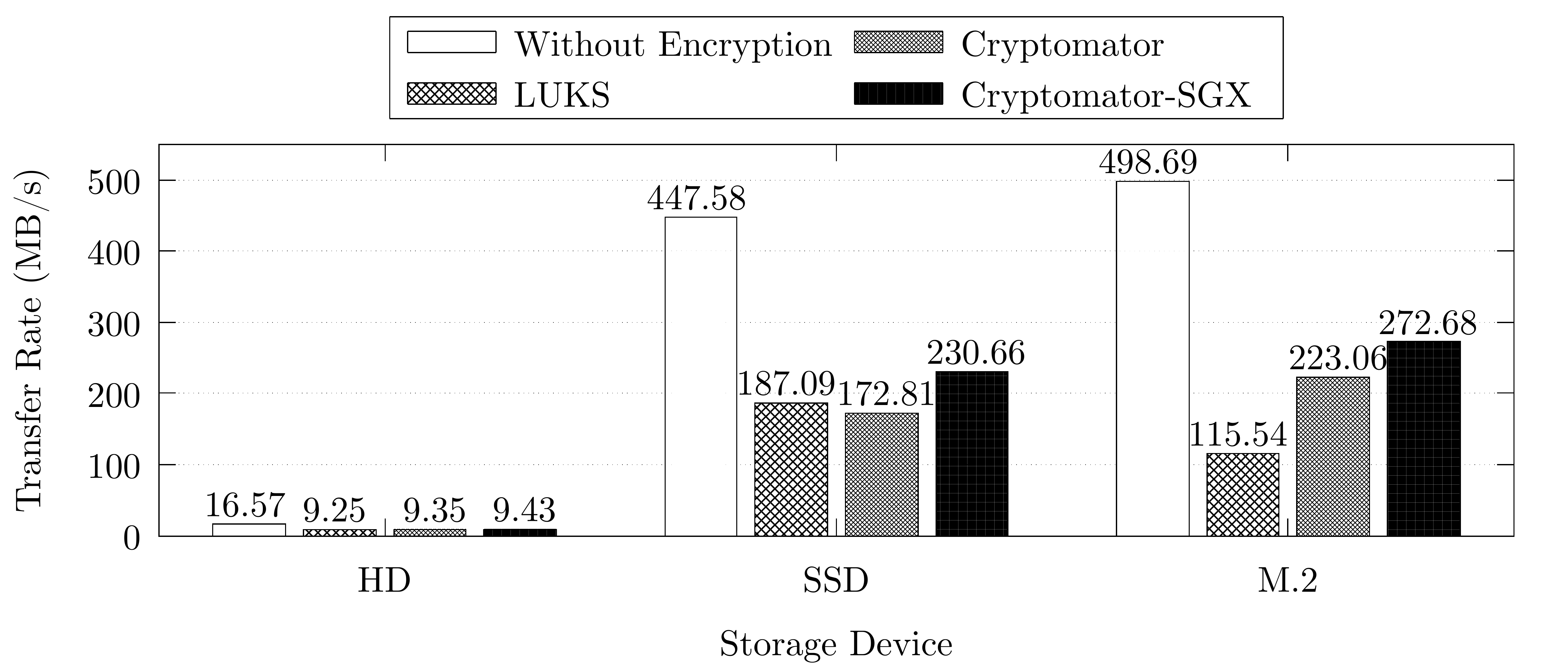}
    \caption{Transfer rate reading a single file with Intel Core i7 9700K CPU.}
    \label{fig:i7_read_iso}
\end{figure*}

Similarly, the results obtained using the Intel Core i7 9700K CPU, for the three storage devices considering a single file read, can be seen in Fig. \ref{fig:i7_read_iso}. In this scenario, the Cryptomator application achieved throughput rates close to or higher than LUKS, with the Cryptomator-SGX solution achieving better performance than both in the three storage devices tested.

\begin{figure*}[ht]
    \centering
    \includegraphics[width=.9\textwidth]{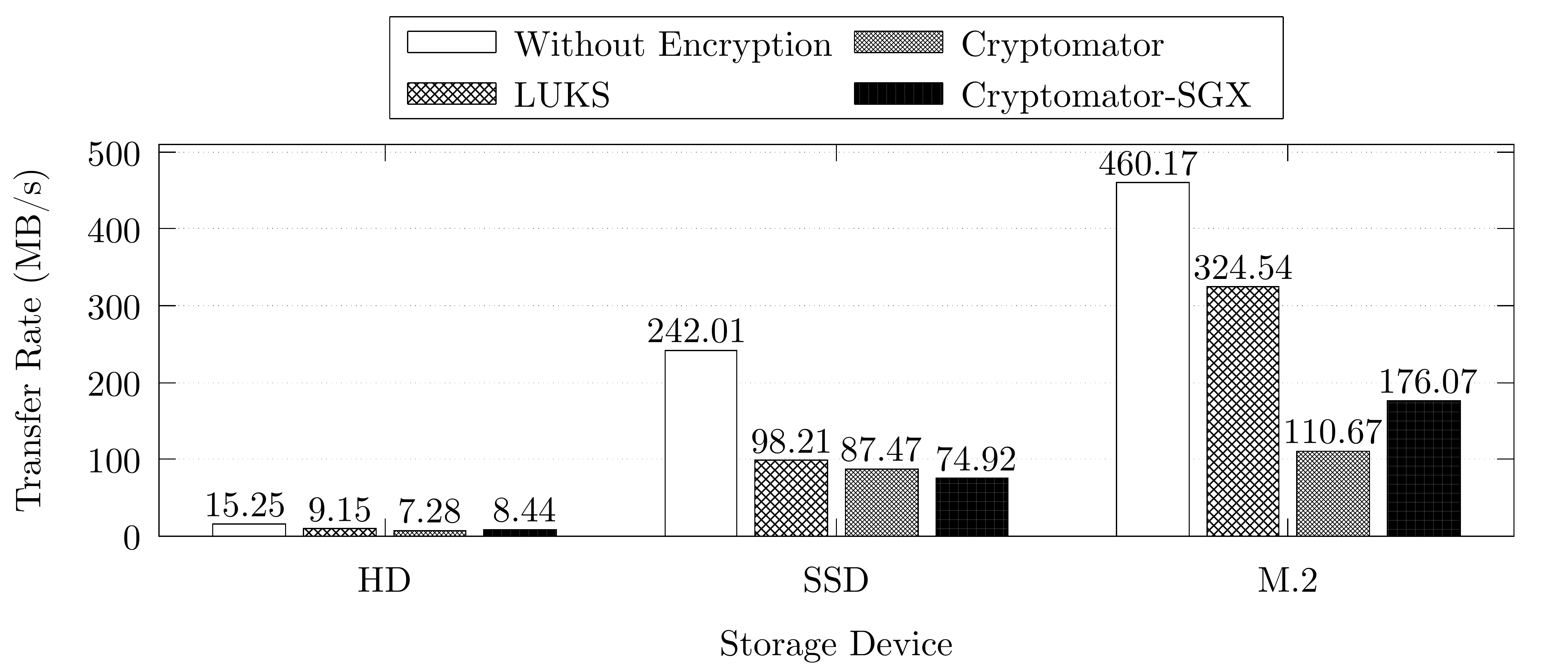}
    \caption{Transfer rate reading multiple files with Intel Pentium G5500 CPU.}
    \label{fig:pentium_read_folder}
\end{figure*}

When considering the reading of multiple and different files, using the Intel Pentium G5500 CPU, the good performance of the proposed solution is also noted, except in the use of the SSD storage device, in which a drop of about 15\% in the transfer rate was obtained, as shown in Fig. \ref{fig:pentium_read_folder}.

\begin{figure*}[ht]
    \centering
    \includegraphics[width=.9\textwidth]{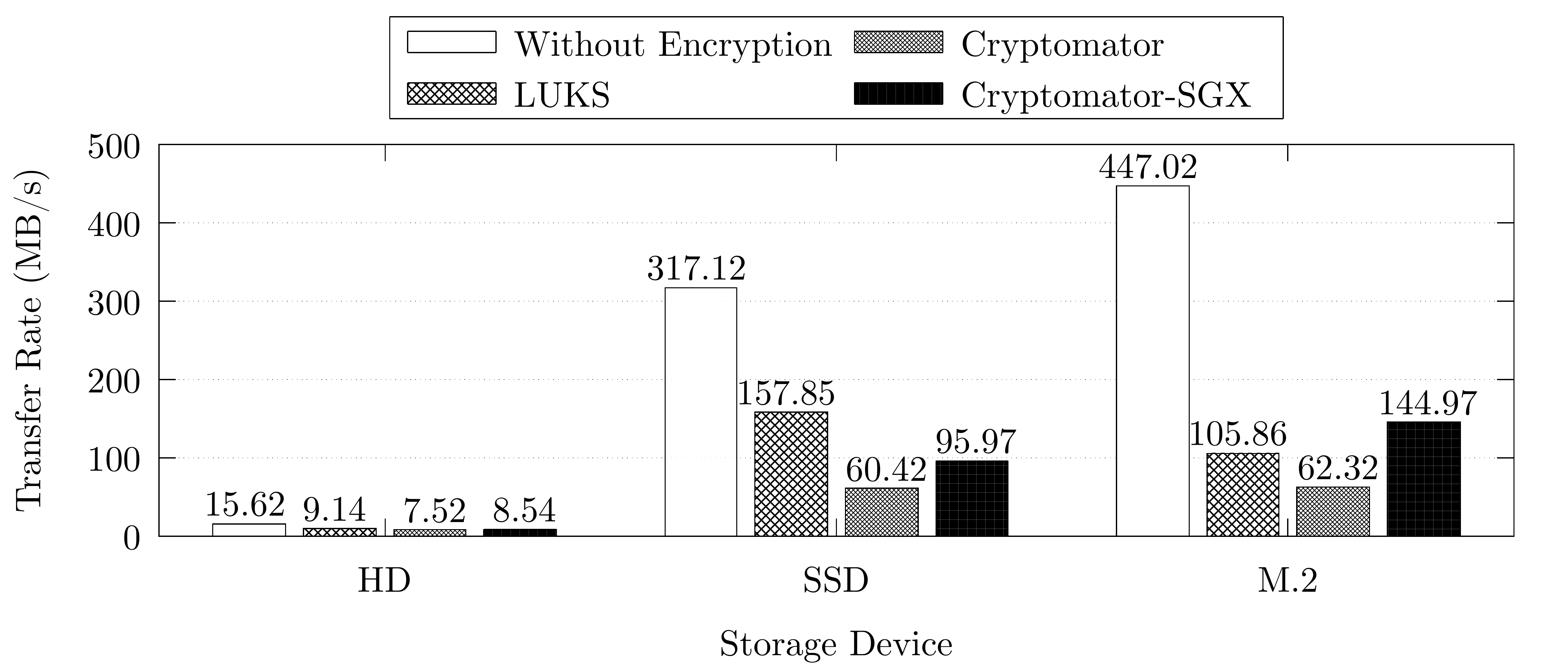}
    \caption{Transfer rate reading multiple files with Intel Core i7 9700K CPU.}
    \label{fig:i7_read_folder}
\end{figure*}

Using the Intel Core i7 9700K CPU, the Cryptomator-SGX solution outperforms the original application on all three storage devices tested, when reading multiple files, as showed in Fig. \ref{fig:i7_read_folder}.

\begin{figure*}[ht]
    \centering
    \includegraphics[width=.9\textwidth]{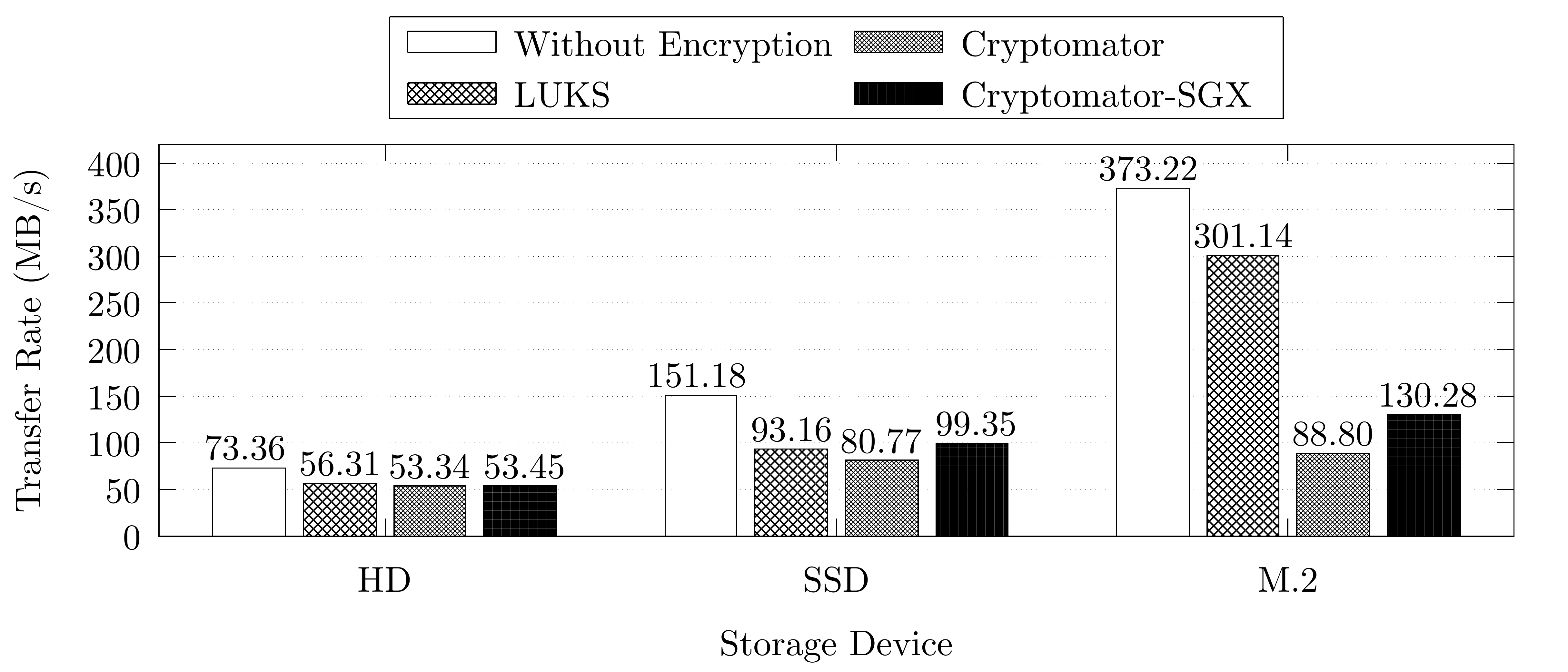}
    \caption{Transfer rate writing a single file with Intel Pentium G5500 CPU.}
    \label{fig:pentium_write_iso}
\end{figure*}

In order to evaluate the Cryptomator-SGX performance on writing operations, we repeat the previous tests, but reading data from RAMDisk and writing in the storage device. The first analysis performed concerns the single file recording using the Intel Pentium G5500 CPU, with results presented in Fig. \ref{fig:pentium_write_iso}. Analyzing these results, we can note that the transfer modes using the multi-threaded capability achieved higher transfer rates, relative to Cryptomator and Cryptomator-SGX, which are limited to using only one thread.

Besides, there is a better performance of the Cryptomator-SGX solution over the original implementation of Cryptomator, as the first uses the native cryptographic functions of the processor as well the execution within the enclave is from native code, written in C++, and optimized by the compiler. Although the data transfer operations through the JNI interface, for the communication with the enclave, and the context changes, generated by the enclave entry and exit, have a considerable computational cost, these two overheads are totally suppressed by the gain obtained in the data sealing operation.

\begin{figure*}[ht]
    \centering
    \includegraphics[width=.9\textwidth]{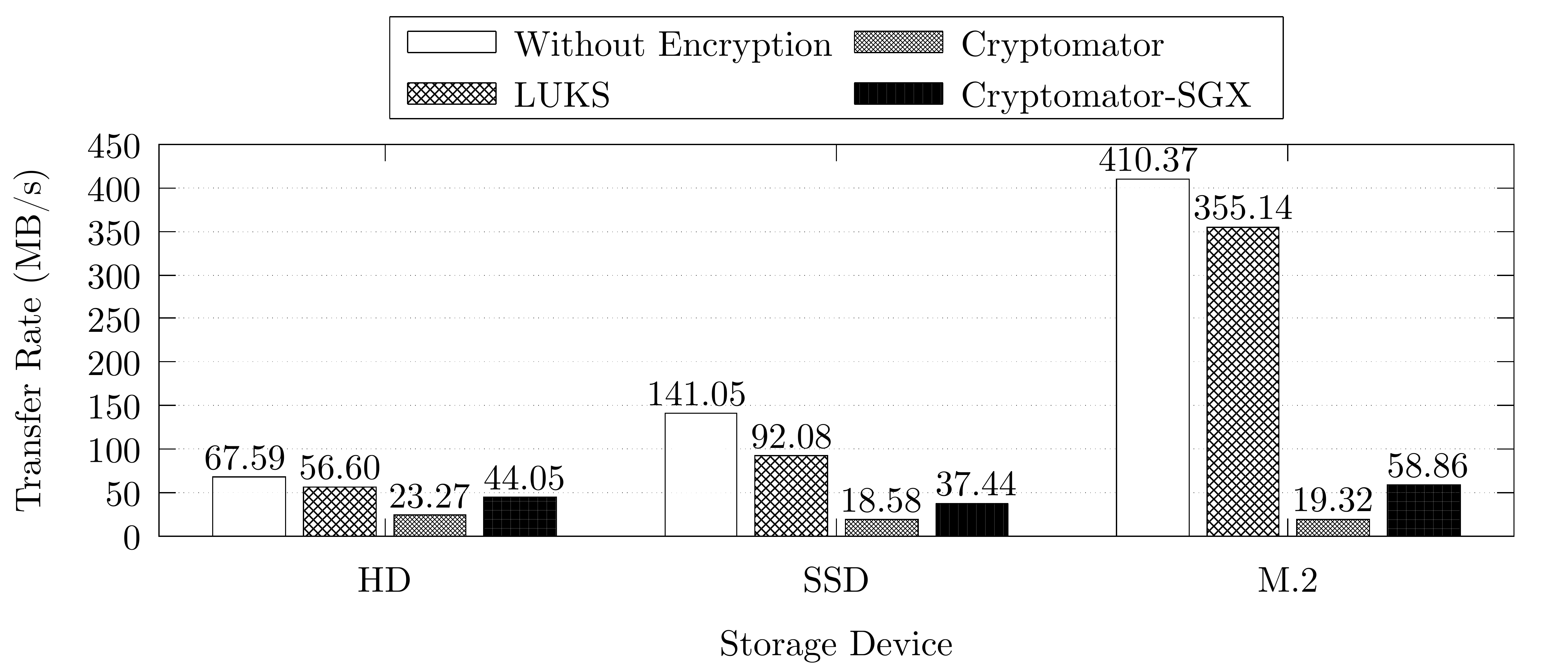}
    \caption{Transfer rate writing a single file with Intel Core i7 9700K CPU.}
    \label{fig:i7_write_iso}
\end{figure*}

Fig. \ref{fig:i7_write_iso} presents the results obtained with the Intel Core i7 9700K CPU in the same scenario: single file writing. As can be seen, Cryptomator and Crypto-mator-SGX perform significantly lower than the other two modes, because they do not utilize the multiprocessing features available on the CPU.

\begin{figure*}[ht]
    \centering
    \includegraphics[width=.9\textwidth]{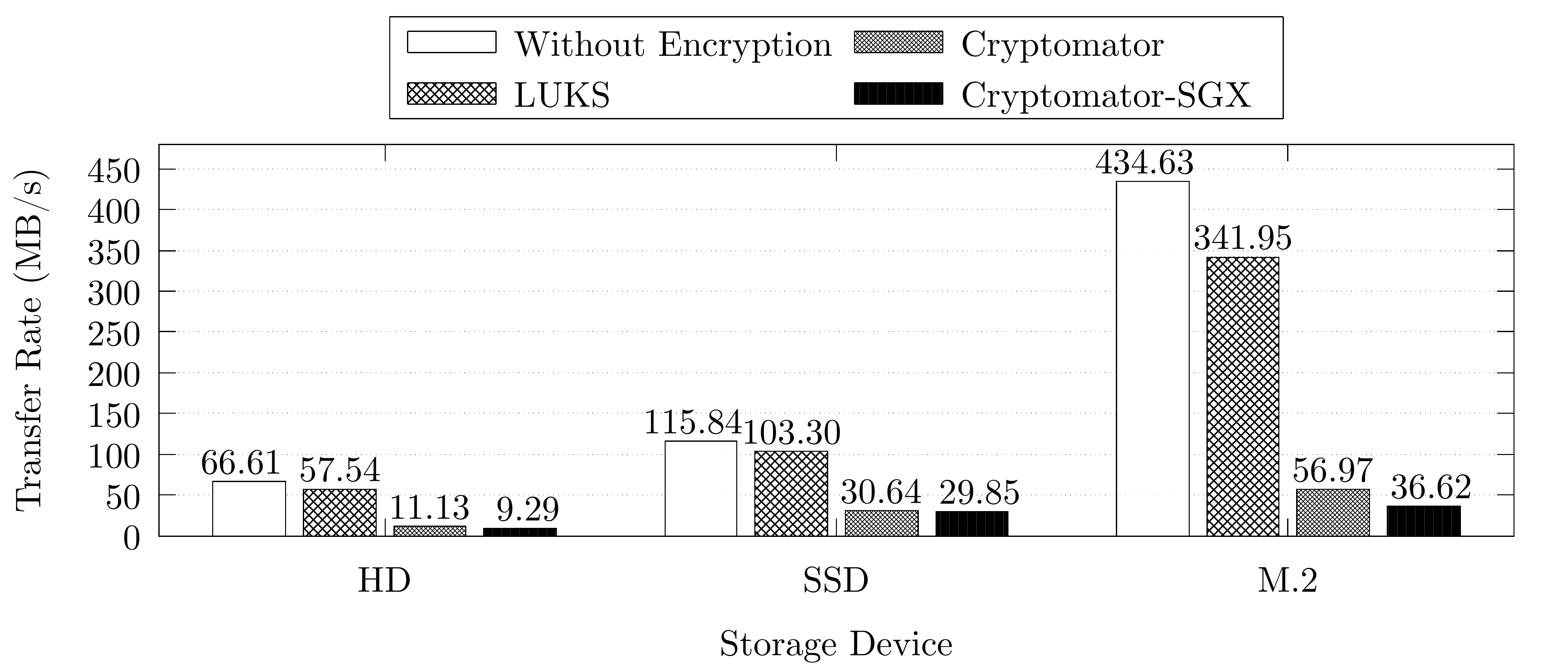}
    \caption{Transfer rate writing multiple files with Intel Pentium G5500 CPU.}
    \label{fig:pentium_write_folder}
\end{figure*}

The second data recording analysis considers several files of different sizes, trying to identify the impact caused by the file system when performing operations to construct the references to such files. For this scenario, the files contained in the previously presented ISO image were extracted, and all contents were copied, as already mentioned. The results for the Intel Pentium G5500 CPU are shown in Fig. \ref{fig:pentium_write_folder}.

Although the Cryptomator-SGX solution presents a substantial performance overhead when using the M.2 storage device, the transfer rates achieved were close to 80\% when compared directly to the original Cryptomator implementation.

\begin{figure*}[ht]
    \centering
    \includegraphics[width=.9\textwidth]{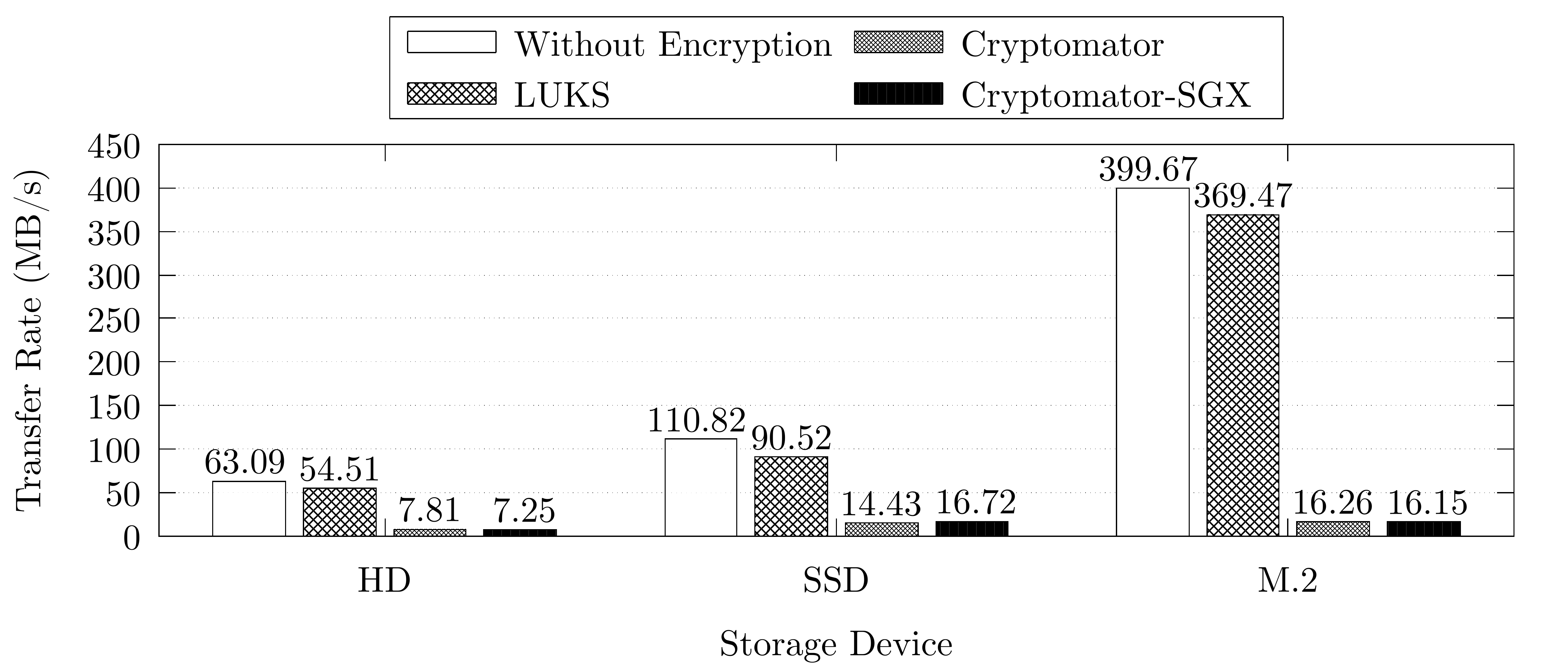}
    \caption{Transfer rate writing multiple files with Intel Core i7 9700K CPU.}
    \label{fig:i7_write_folder}
\end{figure*}

Fig. \ref{fig:i7_write_folder} presents the results obtained using the Intel Core i7 9700K CPU for multiple files writing. In this scenario there is a considerable decrease in performance, both in the original implementation of Cryptomator and in the proposed Cryptomator-SGX, due to the single-thread programming model used by Cryptomator. Nevertheless, the results obtained with the use of Cryptomator-SGX show little difference compared to unmodified Cryptomator.

\section{Security Evaluation}
\label{sec:Security}

Another point that users consider when choosing an application for data encryption is the security it provides and its vulnerabilities. This section is intended to present the threat model, the security analysis of Intel SGX technology and the security analysis of the implemented solution.

\subsection{Threat Model}
\label{sec:ThreatModel}

In the threat model we consider that the adversary aims to access confidential information stored on the user's computer hard disk or cloud storage service, and that he/she has physical access to these platforms for such task. The attacker can even remove the hard disk from user's computer and install it on another machine with greater computing power, in order to apply techniques to find out the password used in key generation, or even the encryption key itself. In addition, it is assumed that the attacker has installed some malicious software on the user's computer in order to obtain the encryption key through a memory dumping, or to obtain the password used to open the container by capturing the data entered into the keyboard through a keylogger.

It is considered that Intel SGX technology works properly and according to its specifications, and that the proposed solution development environment is reliable.

\subsection{Intel SGX Technology Security Evaluation}
\label{sec:SGXSecurity}

To build a system that is considered secure, the Trusted Computing Base (TCB) should be kept as small as possible in order to reduce the chances of success in an attack. In Intel SGX technology, the TCB is composed of the CPU and its internal elements, such as hardware logic, microcode, registers and cache memory, and some software elements used in remote attestation, such as \emph{quoting enclave}.

In the implemented solution, the responsibility of the data encryption was delegated entirely to the enclave. Thus, all warranties provided by the Intel SGX technology are in use during data encryption. Some of these guarantees are:
\begin{itemize}
    \item the cryptographic key never goes outside the processor boundaries;
    \item the memory in which it is running, PRM, is encrypted and has mechanisms against direct memory attacks;
    \item protection against external enclave attacks, even if these attacks come from components with high execution privilege, such as BIOS or hypervisor, as specified in Section \ref{sec:SGX}.
\end{itemize}

However, even though the Intel SGX technology have multiple data protection mechanisms, there are some forms of attacks that it is susceptible to. Intel documentation lists technology limitations, specifying that SGX is unable to prevent side-channel attacks, exploiting cache and data access patterns, or physical attacks against the CPU such as fault injection or reprogramming of machine code functionalities \cite{Brasser2017,Wang2017,Chen2018}.

The Intel SGX technology is also susceptible to the speculative execution attack, called \emph{Spectre} \cite{Kocher2018}. The exploitation of Spectre vulnerabilities to infer secrets contained in an enclave is demonstrated by \cite{Chen2018_2}, where the authors show that an enclave's code execution prediction could be influenced by applications outside the enclave, which could temporarily change enclave control flow to execute instructions that lead to observable cache state changes, and, thus, an adversary can use to learn secrets from the enclave memory or its registers.

One of these problems was addressed by \cite{Ahmad2018}, in which the authors propose a way to reduce the chances of success in a syscall and page fault-based side channel attacks by using an ORAM protocol in conjunction with the Intel SGX. But both side-channel and speculative execution attacks are complex to perform practically.

In addition to the vulnerabilities described by Intel, an SGX enclave may be compromised in cases where the enclave development environment has been compromised, or if the SDK used is not the newest version provided by Intel through its official channel. Enclave security also depends on the developer, since the developer must take precautions when manipulating data within the enclave, thus avoiding problems with data leaks when manipulating pointers or calls outside the enclave.

\subsection{Cryptomator-SGX Solution Security Evaluation}
\label{sec:SolutionSecurity}

In order to validate the security of the implementation, the Intel SGX technology was considered as secure, being validated only the change made in the Cryptomator.

When considering that the Intel SGX technology is secure, it is guaranteed that the data can only be decrypted on the platform where it was encrypted, or if the attacker is able to obtain the decryption key through a brute-force attack. In this scenario, we must ensure that only the data owner has access to them on the platform where they were encrypted, so the password used to open the container has been maintained and is used to encrypt the files name.

In this way, even if an attacker gains physical access to the media and the platform used, he/she will still need the password to access the data. If he/she does not have access to the platform, he/she will need the user's password and also the key used by the enclave to encrypt the data. In a computer memory attack, the key used by the enclave will not be available since it never goes beyond the processor limits. However, with this attack mode, it is possible to obtain the password used by the user to open the container.

This scenario also ensures data confidentiality even if the cloud storage provider is compromised as data sent to it are previously encrypted on the user's local platform. The client-side encryption, allied with the encryption provided by the cloud storage provider, adds an extra layer of security to user data and represents a new obstacle for the attacker to overcome.

Due the Cryptomator software is developed in Java and is separated into three main modules, an attacker can make use of the CryptoLib library and implement its own solution and bypass existing security mechanisms within Cryptomator. Thus, a second security layer is necessary in the CryptoLib, in order to authenticate the request to open the sealed data. This proposal is set out in future work.

\section{Limitations of the Solution}
\label{sec:Limitations}

The solution implemented consists in encrypting only the data of the files that are stored inside the container created through the application. Because of this, some limitations were identified in the application.

Because data sealing adds 560 bytes per encrypted block, since it includes authentication data (AES-GCM), the final file size within the container will always be larger than the original file size. Thus, the deployed solution always use more disk space for the same files over other storage modes. As 32KB size blocks are used, the ratio of file size increase is about 1.7\%.

To perform the encryption process for each file, Cryptomator stores a file with encrypted name inside the container, which corresponds to the source file. In this scenario, the implemented solution could not be used to encrypt the name of the files inside the container by an operating system limitation, as Linux systems impose a maximum filename length of 255 characters for most filesystems and sealing the filename would add 560 bytes to it. To circumvent this limitation, the file name encryption was maintained using the existing AES mode, with the key generated from a password chosen by the user. This limitation eventually added an extra layer of protection in the solution by preventing the opening of the container data, on the platform where data were encrypted, without the user's password being informed.

The limitation that has been addressed in Section \ref{sec:Performance} regarding application performance is due to the fact that it uses only one core for processing. Because of this, only one enclave is used in the process of encryption and decryption of the data, reducing the performance of the system. Another point analyzed was the constant need to convert the data between the Java language and the C++ language through the JNI interface, generating a high cost of processing. Changing the architecture of the solution to make better use of available resources within the system will be the subject of future work. However, the need for conversion between languages can not be removed due to the need to use native code to execute the SGX enclave.

Sealing the data using the sealing key can cause inconvenience, since the data will only be accessible if the application is executed through the processor that was used for sealing them. One way to circumvent this limitation is to create a mechanism within the application for secure data transfer between two platforms on which it is running, allowing remote access and data backup whenever necessary. If the processor used to seal the data is lost in the event of a malfunction or other occurrence, the data cannot be unsealed either, as the sealing key is derived from a unique processor key. This requires mechanisms to securely back up data to be read on another platform or to use another key derivation scheme. Such mechanisms will also be subject of future work.

Using sealing key also makes file sharing difficult through the cloud storage provider. Such a situation can also be circumvented by using remote attestation, or by using another CPU-independent key derivation scheme for file encryption. Finally, the proposed solution, as well as client-side encryption itself, may hinder the application of current data deduplication techniques, being necessary to deploy mechanisms in conjunction with the cloud storage provider, as described in \cite{Yan2018}.

\section{Conclusion and Future Work}
\label{sec:Conclusion}

This work proposed the use of Intel SGX technology for data encryption in an open-source software, to add an extra layer of security to data stored at a cloud storage provider, ensuring data confidentiality even if the storage server leak the data.

From the performance analysis presented in Section \ref{sec:Performance}, we have identified that, in most cases, the implemented solution has achieved superior performance compared to the original application for both data read and write. The increase in performance can be attributed to the fact that the encryption process was carried out directly by the processor, while in the original application it was run through the Java Virtual Machine. When comparing the solution with the multithreading solutions, we have identified that the implemented solution and the original application have much lower performance on high performance media, indicating a limitation on the part of the project implementation.

Considering the security guarantees provided by Intel SGX technology, which are described in Section \ref{sec:SGX}, the developed solution offers an extra level of security by sealing the data using the sealing key in conjunction with the user's password to open the containers. Thus, in an attack on the encrypted data, the attacker will need to discover the user's password, and also the sealing key or gain physical access to the processor used for sealing.

In order to avoid that a user's password can be obtained in a memory attack, it is necessary to change the software so that the password is stored inside the enclave and does not leave its limits. Just as it was possible to seal user file data using Intel SGX technology, it can also seal configuration data. Such change requires adjustments to the structure of the CryptoFS library and may remove compatibility with the main project, and it will be the subject of future work.

Also, it is possible to use a similar approach to \cite{Richter2016}, encrypting the data within the boundaries of the enclave, but using a derived key from an user password, and manipulating that key only within the enclave. Such approach makes the data decryption independent of the processor that encrypted them. Also, the current solution can be changed to use the remote attestation feature and allow container data transfer between two machines running the application over secure channels.

Finally, better performance can be achieved by using all processing cores available on the platform. Such implementation demands a change in the main structure of the application, which treats the requests coming from the operating system, being necessary to add the use of queues and parallel processing, thus allowing one block to be processed by the enclave while another is read or written to the storage device.

The source code of the presented solution is available at \url{https://github.com/utfpr-gprsc/cryptomator-sgx}.

\bibliographystyle{IEEEtran}
\balance
\bibliography{references}

\end{document}